\documentclass{aa}
\usepackage{graphicx}
\begin{document}
\title{Cluster formation versus star formation around six
regions in the Large Magellanic Cloud}
\author{Annapurni Subramaniam}
\institute{Indian Institute of Astrophysics, II Block, Koramangala, Bangalore
560034, India}
\offprints{Annapurni Subramaniam, e-mail: purni@iiap.ernet.in}

\date{Received / accepted}

\authorrunning{Subramaniam}
\titlerunning{cluster formation versus star formation in LMC}

\abstract{
We studied the stellar population 
and star clusters around six regions in the
Large Magellanic Cloud (LMC) in order to understand the correlation between
star formation and cluster formation episodes.
We used the stellar data base of the OGLE II LMC survey and the star
cluster catalogues.
The analysis of the colour-magnitude diagrams
(CMDs) were used to estimate the ages of the stellar population. 
It is found that most of the regions have undergone three
major star formation episodes. The star formation which began 
about 4 Gyr ago, continued upto around 1 Gyr, or continued further.
The other two events have taken place around 300 Myr, and 100 Myr.
A few star clusters were formed during the first of the three star formation events
in 5 regions.  In all the six regions, a good correlation is seen between the star formation
and the cluster formation events which occurred at 300 Myr and 100 Myr. 
The cluster formation events and the fraction of star clusters formed were found to be very 
similar for two regions located to the south-east of the Bar,  suggesting similar cluster 
formation triggers at almost similar instances.
The two recent star formation events seem to  correlate with the 
interactions of LMC with Small Magellanic Cloud (SMC) and Galaxy.
Hence it is 
quite likely that the young populous star clusters in the LMC are formed 
as a result of the star formation events started due to 
galaxy-galaxy interactions and the further propagation of such star formation.
\keywords{Large Magellanic Cloud - star clusters - star formation history}
}

\maketitle

\section{Introduction}
The Large Magellanic Cloud is known to have a very different star formation
history compared to our Galaxy. It is known to house a different variety 
of star clusters, which are not found in our Galaxy. Our Galaxy has very old
and very dense clusters, called globular clusters and a large number of
intermediate to very young clusters with only a few hundred to thousand stars as
members, called open star clusters. In the LMC, only a few globular clusters
are found and they are referred to as red globulars. There are another set
of dense and compact star clusters known as blue globulars, which look like 
globular clusters, but their age ranges from a few giga years to a few million years.
The mechanism responsible for the formation of these dense star clusters in LMC
is not known, when our Galaxy seems to be creating clusters which are hardly bound.

In the recent years, the LMC has been very thoroughly studied using various surveys,
for example, OGLE II (Udalski et al. \cite{u2000}), Magellanic Clouds Photometric Survey 
(Zaritsky et al. \cite{z97}). These surveys were used partly or fully to study the 
star clusters and the stellar population in LMC. However, these two are done separately
such that the spatial correlation between the star and cluster formation 
episodes are not studied.  Girardi et al. \cite{g95} used the star cluster catalogue
of Bica et al. \cite{bea96} to derive the star cluster properties in LMC.
Pietrzynski \& Udalski \cite{pu00} used the OGLE II data and studied the age
distribution of LMC star clusters.
There were also studies on star clusters and stellar population around them,
(for example, Olsen et al. \cite{o98} and Olsen \cite{o99}), but these also do not 
compare the spatial correlation between cluster and star formation episodes. 
It is concluded that in general, the star clusters in LMC are not good tracers
of the stellar population (van den Bergh \cite{v99}). 
This conclusion has been made from the analysis
of the cluster and stellar population in the whole of LMC. 
In this study, an attempt is made
to study the correlation between the star formation and cluster
formation episodes around a few regions in LMC. 

The recent star forming regions, like the 30 Dor and super giant shells found in
LMC indicate
that the star formation which began at one point propagates to larger distances
in the LMC.  The theories put forward to explain these structures include
stochastic self-propagating star formation, SSPSF (Feitzinger et al. \cite{f81}) and
recently by de Boer et al. \cite{b98}, suggesting bow-shock induced star formation.
It is quite likely that these type of groups and star formation processes
existed in older times as well.
If these type of structures existed in LMC, then the cluster population surrounding
a region of stellar population should have some correlation in age. Also, nearby regions
would then show similar cluster formation episodes and similar number of clusters per
episode of cluster formation.  We explore this correlation in this study, by looking
at star clusters in neighbouring regions in LMC.

\section{Data and analysis}
\subsection{Regions}
\begin{figure}
\resizebox{\hsize}{!}{\includegraphics{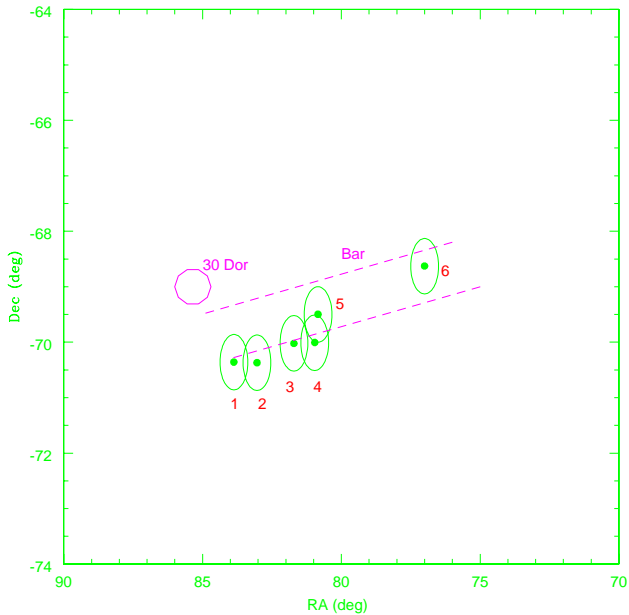}}
\caption{The position of the regions studied here are shown on the LMC. The Bar and
30 Dor are also shown.}
\label{figure1}
\end{figure}
\begin{table*}
\caption{Location of the six regions in the Large Magellanic Cloud.}
\begin{tabular}{lllllll}
\hline
Region &\multicolumn{3}{c}{RA}&\multicolumn{3}{c}{Dec}\\
     & h & m & s & $^\circ$ & $^\prime$ & $^{\prime\prime}$ \\
\hline
Region 1  & 05&35&29.33& $-70$&21&29.39\\
Region 2  & 05&32&09.27& $-70$&22&11.70\\
Region 3  & 05&26&50.33& $-70$&01&23.08\\
Region 4  & 05&23&50.12& $-70$&00&23.50\\
Region 5  & 05&23&21.82& $-69$&29&48.48\\
Region 6  & 05&08&01.10& $-68$&37&37.67\\
\hline
\end{tabular}
\end{table*}
The regions studied here were chosen for a different project, 
that is, to study the stellar population around novae in LMC (see
Subramaniam \& Anupama \cite{sa02}). 
We selected 6 regions for the present analysis.
The stellar data within a radius of a few arcmin is used to study
the star formation
history (SFH) of the region under consideration, whereas the star clusters 
are identified within 30 arcmin radius to study the
cluster formation events.
The regions satisfy the following two conditions - a) there are good
number of star clusters within 30 arcmin radius and b) ages are known 
for most of the identified clusters.
The location of these regions are given in Table 1. These
locations are also plotted in figure ~\ref{figure1}. The dots show the centers
of the regions studied and the big circles around them show the extend of the
region scanned for star clusters.

It can be seen from figure ~\ref{figure1} that the first and the second regions
are located nearby, and similarly the regions third and fourth
are located close by. Of these, regions 1, 2, 3 and 4 are located just outside and south
of the Bar, whereas, region 5 is located within the Bar. The region 6 is located within the
Bar, but towards the north-western side of the Bar.

\subsection{Field stars}
Field stars within a radius of a few arcmin were
identified from the OGLE II survey (Udalski et al. \cite{u2000}). 
As more observations were found to be available in the I passband, we used 
the photometric data in the V and I passbands and V vs (V$-$I) 
colour-magnitude diagrams (CMDs) of the identified field stars were used in
further analyses.  

The stellar data are 
corrected for an assumed LMC reddening of $E(V-I)= 0.10$ mag.
Although the reddening in the LMC has been shown to be varying and
clumpy (Udalski et al. \cite{u99}), the uniform reddening assumed here was found to
be satisfactory while fitting the isochrones to the main-sequence
of the CMDs.
Assuming the relation $A_V=3.24 E(B-V)$ and $E(V-I)=1.37 E(B-V)$, the
value of $A_V=0.355$ mag. This value of extinction agrees (within errors) with 
that estimated by Dolphin \cite{d00}. Following Pietrzynski \& Udalski \cite{pu00} a 
distance modulus of 18.24 mag for the LMC is assumed here. For this value of
distance modulus, 1 arcmin corresponds to 13.4 pc on LMC. 

The different stellar populations are identified by
isochrone fits to the CMDs. An isochrone fit to the MS identifies the youngest
population while the isochrone fits to the RGB identifies the older population.
The slope of the RGB and the brightest point of the RGB are dependant on the 
age and metallicity of the stellar population. The analysis done in Subramaniam
\& Anupama \cite{sa02} found that the metallicity of the stellar population
agrees well with the Z=0.008 isochrones.  Hence we adopt the value Z=0.008
for the metallicity.  The isochrones were obtained from Bertelli et al. \cite{bea94}.

The limiting magnitude in the OGLE II data is around $V = 21.0$~mag. This implies
the stars in the MS are younger than about 1.6 Gyr, while the RGB stars are
a mixture of both young and old population. Therefore the present data is not suitable
to understand the star formation history older than around 4 Gyr.

\subsection{Star clusters}

The star clusters in the vicinity of six regions were identified and their properties
obtained based on the
following catalogues: Pietrzynski et al. \cite{pea99} (P99), Bica et al.\
\cite{bea99} (B99), Bica et al.\ \cite{bea96} (B96), Pietrzynski and Udalski 
\cite{pu00} (PU2000). 
B96 presented integrated UBV photometry of 624 star clusters and associations 
in the LMC. They estimated the ages of the clusters based on their integrated 
colours and hence classified the clusters into SWB types (Searle, Wilkinson \&
Bagnoulo \cite{swb80}), which is basically an age sequence. This classification can
be used to obtain the approximate age of the clusters.
B99 is a revised version of the above catalogue and contains about 1808 star
clusters for which the positions and extents are tabulated. 
Pietrzynski et al.\ \cite{pea99} presented photometric data of 745
star clusters and their nearby field, of which 126 are new findings. 
Pietrzynski and Udalski \cite{pu00} estimated the ages for 600 star clusters
presented in the P99 catalogue. The catalogues in B99 and P99 were used to
identify the clusters, while B96 and PU2000 were used to estimate the ages of 
the identified star clusters.

Clusters have been identified within 30~arcmin radius around 6 regions using 
B99 around the regions tabulated in table 1.
109 clusters have been identified near 6 regions. Of these,
age estimates for 89 clusters could be obtained from PU2000 and B96. 
The B96 gives the age of the cluster in terms of groups. As the interest is
in the overall age of the cluster population rather than the ages of the
individual clusters, age groups give a better insight. Therefore, even those
clusters whose exact age is known are also grouped. The number of clusters 
detected near each region, the number for which the age is known and the number 
of clusters in various age groups are tabulated in Table 2. The clusters are
grouped into the following five age groups: 
\begin{description}
\item (a) clusters with ages $\log \tau < 7.5 $, indicating clusters 
of very young age
\item (b) clusters with ages $7.5 \le \log \tau < 8.0$, indicating clusters
which are relatively young. 
\item (c) clusters with ages $8.0 \le \log \tau < 8.5$,  
equal to 8.0 and less than 8.5, indicating a group of moderately young clusters.
\item (d) clusters with ages $8.5 \le \log \tau < 9.0$, 
\item (e) clusters with ages $\log \tau \ge 9.0$, indicative of intermediate to old population.
\end{description}

Field stars within a radius of a few arcmin 
are analysed to study the star formation history, while clusters within 30 
arcmin ($\sim$ 400 pc) radius are considered to identify the cluster formation
episodes.
The choice of larger radius for the clusters is justified as they are being used to study
the events which took place on relatively larger
scales. It is found that
the supergiant shell LMC 4 is about 1 Kpc in diameter.  Therefore, we have chosen
very similar length scale for identifying the star clusters.
\begin{table*}
\caption{Statistics of star clusters identified near the regions.}
\vspace{0.2cm}
\begin{tabular}{cccrrrrr}
\hline
Region &  No. of clusters & No. of clusters & \multicolumn{5}{c}{Age groups}\\
     &  within 30 arcmin&  with age known & $\le$ 7.5&7.5\,--\,8.0&8.0\,--\,8.5
&8.5\,--\,9.0&$\ge$ 9.0\\
\hline
Region 1&     17 &15 & 1 & 1 & 6 & 5 & 2 \\
Region 2&     15 &12 & 1 & - & 7 & 3 & 1 \\
Region 3&     27 &22 & - & 4 &11 & 7 & - \\
Region 4&     24 &19 & - & 2 &11 & 4 & 2 \\
Region 5&     17 &13 & - & 3 & 3 & 3 & 4 \\
Region 6&      9 & 8 & - & 4 & 1 & 2 & 1 \\
\hline
\end{tabular}
\end{table*}

\begin{figure*}
\centering
\includegraphics[width=17cm]{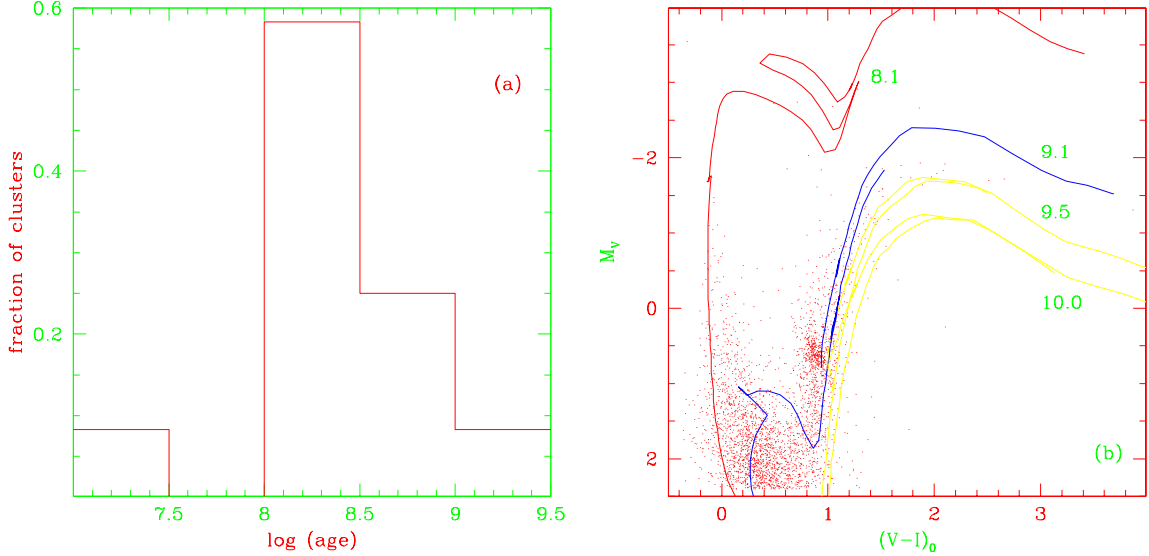}
\caption{Left panel shows the histogram of the normalised cluster fraction
around region 1, against age. Right panel shows the CMD of 
4238 stars within 2 arcmin
radius. The isochrones are plotted on the CMD and their log (age) are indicated.}
\label{figure2}
\end{figure*}

\section{Results}
\subsection{Region 1}
Within a radius of 400 pc, we identified 17 star clusters around this region and ages 
are known for 15 of them. We find that,
forty percent of the clusters have ages in the range 100\,--\,300 Myr, 33.3\% of clusters
have ages in the range 300\,--\,1 Gyr and 13.3\% clusters are older then 1 Gyr. Hence, this
region experienced a 2.5 times enhancement in the formation of clusters, between 
300\,--\,1 Gyr, and increased a little more around 300 Myr and continued till 
100 Myr. In the last 100 Myr, this region has formed 1 cluster older than 30 Myr 
and one cluster younger than 30 Myr. Hence there has been a constant formation 
of clusters with varying rates. Figure~\ref{figure2} shows the histogram of 
the normalised fraction of clusters with respect to age. 

The CMD and LF of 4238 field
stars within a radius of 2 arcmin is shown in figure~\ref{figure2}.
A broad RGB, a prominent clump and a few bright giants are seen 
in the evolved part of the CMD. The broad RGB is indicative of stars with a 
range in ages. The isochrone fit to the right most end of the RGB indicates that 
the reddest stars are 2.5 Gyr old. Stars belonging to a population older than
2.5 Gyr is not found. It is also seen that at the fainter end of the CMD, the MS 
and the RGB are not well separated, but joined together due to the presence of 
sub giants and the isochrone fits show that the
stellar population are of ages 1.3 Gyr and 1 Gyr. The
MS is wide upto $M_V = 0.5$ mag and the width decreases at brighter magnitudes.
This probably indicates a decrease in star formation. The isochrone of age
300 Myr shows that the star formation probably continued till then. 
The vertical extension of the RGC, as seen in the CMD also indicates 
that stars younger than 1 Gyr, in the range 300\,--\,1 Gyr,  are present. 
The well populated MS also indicates the same. The brightest part of the CMD is 
found to be 100 Myr old. There is no indication of the presence of stars younger 
than 100 Myr. 

It is seen  that the star formation event between 1.0 and 2.5 Gyr 
resulted in the formation of two
star clusters. The star formation which continued to younger ages,
 resulted in the formation of 5
star clusters, which fall in the age range 300 \,--\, 1 Gyr. 
This event probably ended around 300 Myr with the creation of another 6 star
clusters. The above two are seen as 
enhancements in the
cluster formation between 300 Myr \,--\, 1 Gyr and 100 Myr \,--\, 300 yr.
The latest star formation event at 100 Myr resulted in one cluster formation.
We do not see stars less than 100 Myr old, though one cluster is found which is aged less
than 30 Myr. The cluster and star formation events are correlated, except for the youngest
population of clusters.

\begin{figure*}
\centering
\includegraphics[width=17cm]{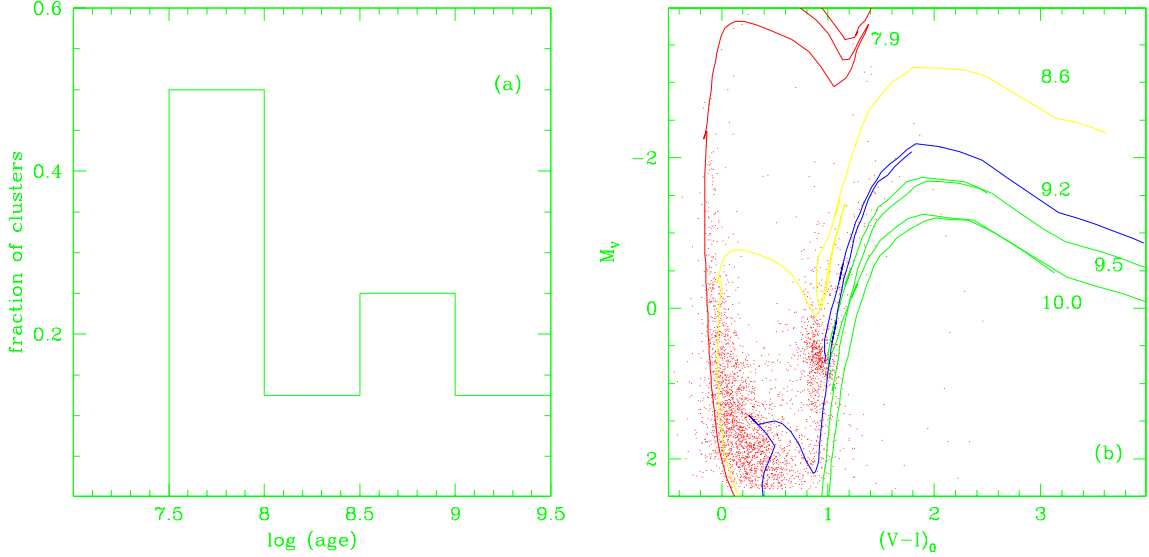}
\caption{Left panel shows the histogram of the normalised cluster fraction
around region 2, against age. Right panel shows the CMD of 3018 stars within 3 arcmin
radius. The isochrones are plotted on the CMD and their log (age) are indicated.}
\label{figure3}
\end{figure*}
\subsection{Region 2}
15 star 
clusters were identified within a radius of 30 arcmin ($\sim$ 400 pc), 
of which the ages are known for 12.
58\% of the clusters in this region have ages between 100\,--\,300 Myr, 25\% 
between 300 Myr and 1 Gyr. Only 8\% of clusters are either younger than 30 Myr or
older than 1 Gyr. Hence the bulk of cluster formation has occurred in the 
100-300 Myr range with a tapering towards 1 Gyr. 
A histogram of the normalised cluster fraction against age is shown 
in Figure~\ref{figure3}.

The field star population within 3 arcmin (40 pc) radius 
is studied based on a CMD of 3018 stars. The CMD is shown in figure~\ref{figure3}.
Isochrones corresponding to 10 Gyr, 3.2 Gyr, 1.3 Gyr and 125 Myr are
used to fit the CMD and estimate the ages of the stellar population. 
A few stars belonging to 
the 10 Gyr population are seen in the RGB, as indicated by the 10 Gyr isochrone. 
There seems to be no stellar population between 10 Gyr and 3.2 Gyr. 
Isochrones of ages 3.2 Gyr and 1.3 Gyr 
fit the RGB and the RGC as well as the sub giant branch stars which connect the 
MS and the RGB at the fainter end of the CMD. This implies that a bulk of the
stars were formed during this time. The stars towards the right side of the 
1.3 Gyr isochrone at the RGB belong to a younger population. 
The MS between $M_V$ 1.0\,--\,$-$1.0 shows scatter, indicating continued star
formation, probably on a reduced scale.
The brightest stars in the MS are about 125 Myr old. 

A small enhancenment in the star cluster formation occured between 300 Myr \,--\, 1 Gyr
and a substantial enhancement occured between 100 Myr \,--\, 300Myr. When we compare
these with the star formation episodes,
we find that corresponding to the star formation between 3.2\,--\,1.3 Gyr, there
is one cluster formed in the region. The continued star formation created 3 more clusters.
Corresponding to the star formation event at 125 Myr, we see maximum number of star 
clusters in the range 100\,--\,300 Myr. There is one cluster which is less than 30 Myr old,
but no field population is found younger than 125 Myr. 
Thus the star formation and the cluster formation are correlated, except for the
youngest cluster formation event. This is similar to what is observed in region 1.

\begin{figure*}
\centering
\includegraphics[width=17cm]{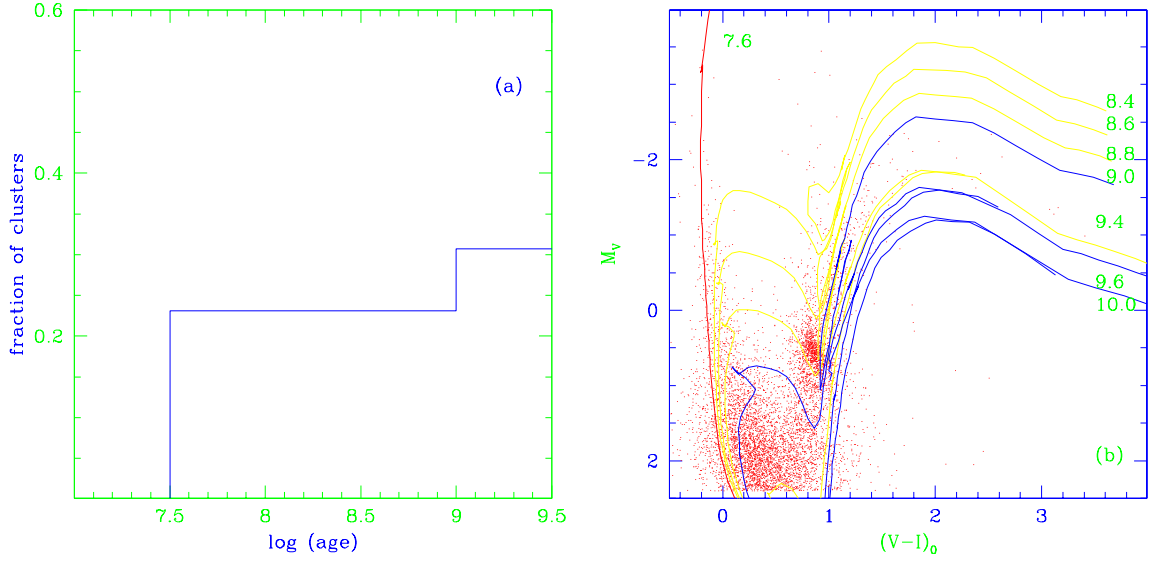}
\caption{Left panel shows the histogram of the normalised cluster fraction
around region 3, against age. Right panel shows the CMD of 2152 stars within 3 arcmin
radius. The isochrones are plotted on the CMD and their log (age) are indicated.}
\label{figure4}
\end{figure*}
\subsection{Region 3}
27 star clusters 
are found within the 30 arcmin radius, of which the ages are known for 22. 
18\% of the clusters have ages within 30\,--\,100 Myr, 50\%  have ages between 
100\,--\,300 Myr and 32\% have ages between 300\,--\,1 Gyr. These indicate that
there was a burst of cluster formation some time during 100\,--\,300 Myr with a 
tapering star formation before and after this burst. The normalised cluster 
fraction is shown against age in figure~\ref{figure4}. 

The CMD of 2152 stars within a radius of 3 arcmin is shown in 
Figure \ref{figure4}. The CMD shows a scattered RGB and MS, and a RGC.
The isochrone fits reveal that the oldest traceable population is about 4.0 Gyr. 
The star formation then continued to about 1.3 Gyr. This is indicated by the fit 
of the isochrone to the subgiant branch location and the red giants of the RGB. 
It is clearly seen that the clump population basically belongs to the 1.3 Gyr 
population and younger. A few stars can be seen to the
left of the 1.3 Gyr isochrone at the RGB, which indicates that the star formation 
continued for more recent times, probably with less vigour. 
The broad MS and the subgiant branch  located in between 
the MS and the RGB, support the above fact. There is a clear signature of the 300 Myr
old population as indicated by the isochrone. There are only a few stars which are
birghter than the turn off of the 300 Myr isochrone in the MS, and these are 
found to be 100 Myr old.  

In this region, we see that the cluster formation experienced an enhancement between
100 Myr \,--\, 300 Myr, with a few clusters being formed before and after that.
The field star population do not show a strong presence of RGB population, especially when
compared to the CMDs of regions 1 and 2.
This indicates that the star formation event in the age range 1.3\,--\,4.0 Gyr
was not very strong. This star formation event did not produce any star clusters also.
The star formation seems to have continued upto 300 Myr, where we see that most of
the star clusters are formed. The star formation which occured at 100 Myr has resulted
in the formation of 4 star clusters. The cluster and star formation events are thus
correlated in this region also.

\begin{figure*}
\centering
\includegraphics[width=17cm]{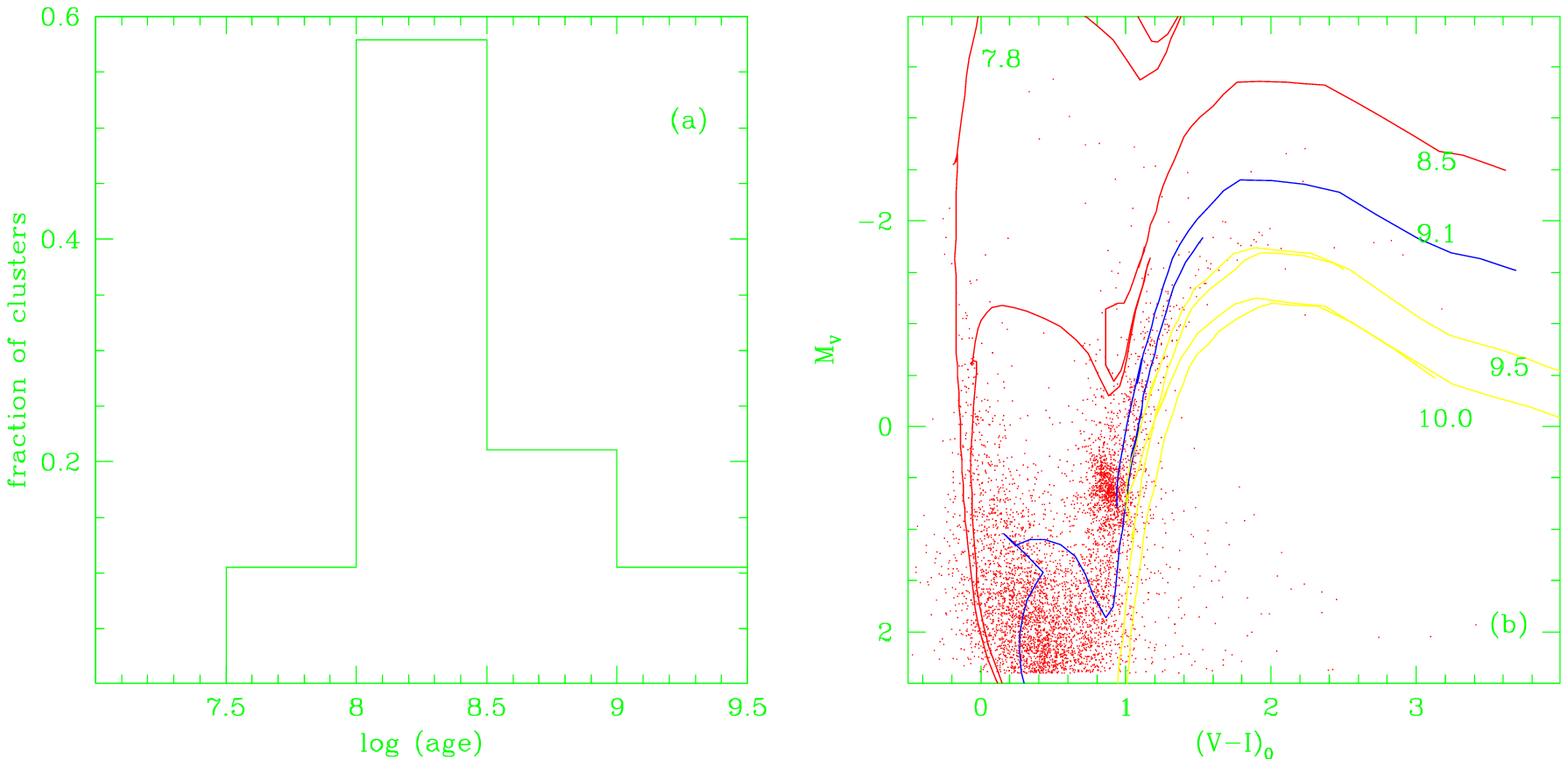}
\caption{Left panel shows the histogram of the normalised cluster fraction
around region 4, 
against age. Right panel shows the CMD os 5244 stars within 2 arcmin
radius. The isochrones are plotted on the CMD and their log (age) are indicated.}
\label{figure5}
\end{figure*}
\subsection{Region 4}
There are 24 star clusters 
in the vicinity and ages are known for 19 of them. 
 The maximum number of star clusters were 
formed during the period 100\,--\,300 Myr, when 58\% of the clusters were formed. 
21\% of clusters were formed during 300\,--\,1 Gyr period, 10\% formed during 
the 30\,--\,100Myr period and the rest during the period before 1 Gyr. It appears
that this region has been forming clusters from the beginning, with an increase
in the rate before 1 Gyr, which increased further between 300\,--\,100Myr.
Subsequently the rate appears to have and then reduced and stopped by 30 Myr. 
Figure \ref{figure5} shows the normalised fraction of the clusters. 

The CMD of stars within 2 arcmin radius plotted in Figure \ref{figure5}
comprises of 5244 stars. The RGB is well populated with a prominent RGC. The 
isochrone fit reveals that the the stars in the RGB have ages in the range 
1.3\,--\,3.2 Gyr. Very few stars belonging to the age of 10 Gyr is seen in the CMD.
The broad MS below $M_V$ = 0.0 mag indicates a continued star formation from about
1.3 Gyr to lesser ages. This probably continued till 300 Myr, as indicated by the
isochrone fit.  The brightest part of the MS is found to be 63 Myr old.  

The cluster formation episode is seen to be quite strong in the age range 
100 Myr \,--\, 300 yr, when a substantial increament in the clusters formed in seen.
On comparing the star and cluster formation episodes, we find that
the star formation event which occured in the age range 3.2\,--\,1.3 Gyr has
managed to form two star clusters. The star formation which continued to younger
times resulted in the formation of 4 star clusters. The star formation 300 Myr
coincides with the cluster formation around the same time, resulting in the formation
on 11 star clusters. The latest star formation event at 63 Myr, resulted in the formation
of 2 star clusters. These indicate a good correlation between star and cluster formation
events.

\begin{figure*}
\centering
\includegraphics[width=17cm]{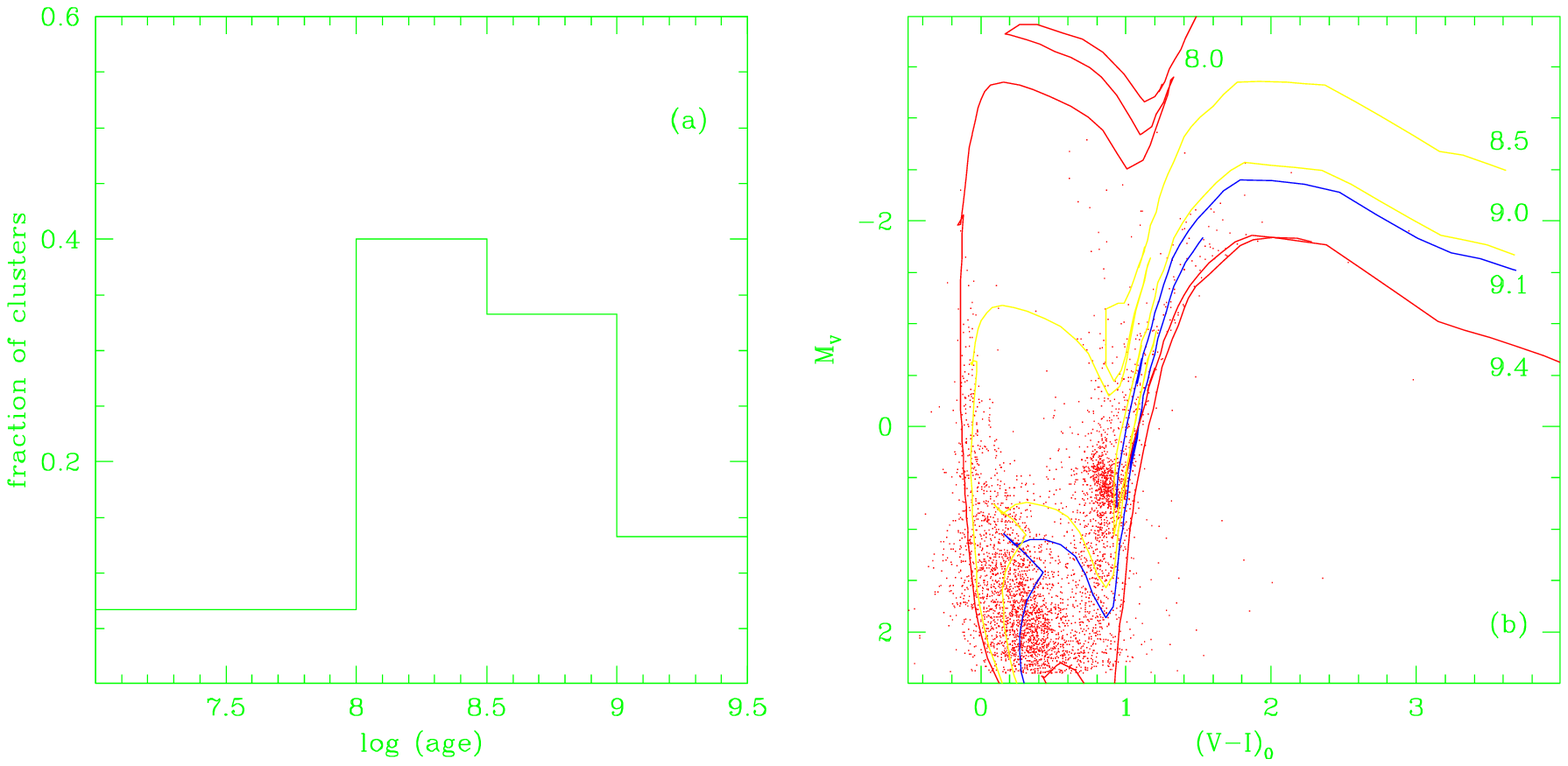}
\caption{Left panel shows the histogram of the normalised cluster fraction
around region 5,
 against age. Right panel shows the CMD os 6503 stars within 2 arcmin
radius. The isochrones are plotted on the CMD and their log (age) are indicated.}
\label{figure6}
\end{figure*}

\subsection{Region 5}
This regions is situated right at the center of the LMC Bar.
Seventeen clusters are found in the vicinity, of which the ages are
known for 13 clusters. Cluster formation in this regions appears to be at a more
or less constant rate until 30 Myr.
31\% of clusters are with ages more than 1 Gyr, and 23\% of clusters
in the lower age ranges. Hence the cluster formation rate was lowered around 
1 Gyr, after an initial higher rate. The normalised cluster fraction is plotted against
age in figure \ref{figure6}

The CMD of 6503 field stars within a region of 2 arcmin (27 pc) radius is
plotted in Figure~\ref{figure6}.
The RGB, RGC and the MS are well populated. The isochrones corresponding to the ages
10 Gyr, 4 Gyr, 2.5 Gyr, 1 Gyr, 630 Gyr, 400 Gyr, 251 Gyr and 40 Gyr are plotted on 
the CMD. These isochrones seem to fit most of the
stars in the RGB. It can be clearly seen that this region has experienced a more or less
continuous star formation from 4 Gyr. We do see some stars belonging to 
the 10 Gyr population.
Most of the RGB stars are located between the isochrones of ages 4 Gyr and 250 Myr.
The isochrone 
fit to the MS of the CMD shows that the brightest MS stars are 40 Myr old. 

The 
field stars convey almost the same information as the star clusters, regarding 
the star formation history of this region. Both indicate a constant star
formation in this region from 4 Gyr till 30-40 Myr ago. The star cluster
formation is also seen to be continuous and without much fluctuation over this
period.
 We see that in this region, the star and the cluster formation episodes are
very well correlated.

\begin{figure*}
\centering
\includegraphics[width=17cm]{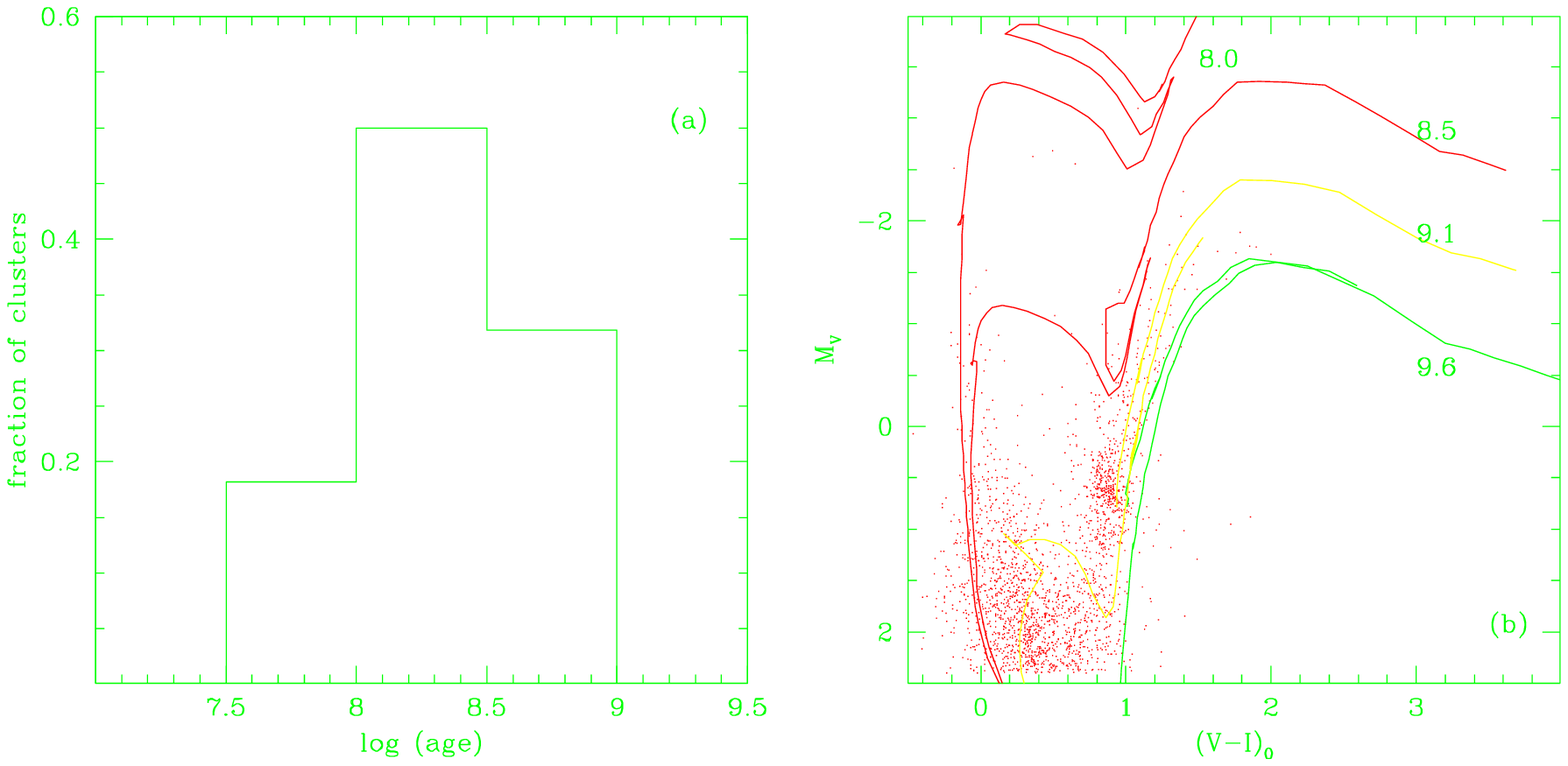}
\caption{Left panel shows the histogram of the normalised cluster fraction
around region 6,
against age. Right panel shows the CMD os 3302 stars within 4 arcmin
radius. The isochrones are plotted on the CMD and their log (age) are indicated.}
\label{figure7}
\end{figure*}
\subsection{Region 6}
In this region, 9 star clusters have been identified 
within 30 arcmin radius and the ages of 
8 clusters are known. 50\% of the clusters have ages in the range 30\,--\,100 Myr, 
12.5\% have ages in the range of 100\,--\,300 Myr, 25\% in the range 
300\,--\,1 Gyr and 12.5\% have ages beyond 1 Gyr. This shows that there has been 
a constant cluster formation, with a rate which doubled during the 300\,--\,1~Gyr 
period and quadrupled during 30\,--\,100 Myr period. 
A histogram of the normalised cluster fraction against age is shown in figure \ref{figure7}.

3302 field stars located within 4 arcmin (54 pc) radius were identified and the
CMD of the field stars are plotted in Figure~\ref{figure7}. 
The CMD shows a very well 
populated MS with a fair number of young stars. The RGB is well populated with a 
slightly scattered RGC.
The isochrones corresponding to the ages 10 Gyr, 3.2 Gyr, 1.6 Gyr, 400 Myr and
79 Myr are shown in the CMD. The 3.2 Gyr and 1.6 Gyr isochrones fit the bulk of
the RGB stars. It can be seen that the 1.6 Gyr isochrone fits the subgiant 
branch as well.
A few brighter subgiant branches are seen, indicative of younger stars.
The CMD shows a population of stars directly above the RGC, which is like the extension 
of the RGC. These can be fitted with isochrones of age 400 Myr. This is indicative 
of a star formation event around 400 Myr. The
CMD of this region shows an almost detached MS and the RGB at the magnitude level,
$M_V = 1.5$ mag, and only a few subgiant branches are seen connecting the two at 
brighter magnitudes. This indicates that the RGB population is relatively older.  
The brightest stars in the MS of the CMD is fitted very well with the 79 Myr 
isochrone, indicating that 
the star formation stopped around 79 Myr ago. This youngest population is also
found to have a good number of stars indicating a relatively stronger star formation
event. 

The star formation event duting the age range 3.2\,--\,1.6 Gyr has produced one cluster
in the vicinity. The star formation which occured at 400 Myr has probably resulted in the
creation of two star clusters. This is seen as a small enhancement in the star cluster
formation in figure~\ref{figure7}. We do not see any significant star formation event 
between 400 and 79 Myr, though we see one cluster in this age range. The field stars
show a strong star formation event at 79 Myr, a corresponding cluster formation event
is also seen resulting in the formation of 4 star clusters. This is seen as a substantial
enhancement in the cluster formation between 30 \,--\, 100 Myr period. 
Therefore, we see that
the  star formation and the cluster formation events are more or less correlated.

\section{Discussion}
\subsection{star and cluster formation events}
\begin{table*}
\caption{Age estimates of the field stellar population}
\begin{tabular}{lcccc}
\hline
Region & old  & intermediate & moderately & young  \\
     &        &             &  young       &        \\
     &  (Gyr)      &  (Gyr)      &  (Myr)       & (Myr) \\
\hline
Region 1 & - & 2.5\,--\,1.0 & 300 & 100 \\
Region 2 & 10 & 3.2\,--\,1.3 &  - & 125  \\
Region 3 & - & 4.0\,--\,1.3 & 300 & 100  \\
Region 4 & 10 & 3.2\,--\,1.3 & 300 & 63 \\
Region 5 & 10 &\multicolumn{2}{c}{ 4.0\,--\,250} & 40 \\
Region 6 & 10 & 3.2\,--\,1.6 & 400 & 79         \\
\hline
\end{tabular}
\end{table*} 
The results presented in the last section is summarised in table 3.
It can be seen from table 3 that the major star formation events
which occurred in the regions studied are mostly three or less than  that.
These events in some regions have continued and hence produced stars for
longer duration.  We consider the star formation events one by one below.

We do not attempt to study the correlation between the cluster and star
formation events before 4-5 Gyr, as the data used here are unsuitable
for this purpose. Also, it is well known that LMC does not have star
clusters in the age range 3\,--\, 10 Gyr. There are only a few star
clusters which are older than 10 Gyr and we do not study the star formation
events in this age range. We see from the previous section that, the density of
the very old population is very small and in some regions, no trace of this
population is found.

The major event of star formation started somewhere around a few Gyr in most
regions of the LMC. In the regions studied here, we find that
the intermediate age stars are formed from 4 Gyr , where 4 Gyr is the
upper age limit. This star formation event has started at three different times,
4 Gyr, 3.2 Gyr and 2.5 Gyr. This event later seems to have ended or decreased in
intensity by 1.0\,--\,1.6 Gyr. 
The star cluster population in this age range is present in 5 of the 6 regions studied.
In the sixth region, we do not find much of the intermediate age stars, which is well
correlated with the absence of any clusters. 
In most of the regions the star formation continued to more recent times, but in
a subdued manner.  The continued star formation to ages
less than 1 Gyr is seen in the CMDs and this corresponds to the continued cluster
formation events.

The moderately young population of 250\,--\,400 Myr is seen in 5 of the 6 regions
studied here. In all these regions, an enhancement in the fraction of clusters is
also observed. In the case of region 5, the star formation is continued till 250 Myr,
which is also reflected in the number of clusters found in this age range.
In the case of region 2, the 300 Myr episode is not seen. Stars belonging to 125 Myr
old population is seen in the CMD of region 2, and a cluster formation event has
occured between 100 and 300 Myr. So we assume that the star and the cluster
formation events coincide.

The most recent star formation in the regions studied here is found to be around 100 Myr,
except for region 5. Star clusters also show an episode of cluster formation 
around this time. In the case of region 2, the cluster formation episode has
occurred at about 125 Myr, and for region 5, it has occurred at about 40 Myr.
In two regions one cluster each have been found, with ages less than 30 Myr, the
corresponding star formation event is not seen in the CMDs. This probably would
indicate that this latest star formation event, which resulted within the last
30 Myr is probably an isolated event, producing very less number of field stars.

Girardi et al. \cite{g95} estimated the age distribution of star clusters in LMC.
They found three periods of enhancement in the formation of star clusters in LMC,
namely at 0.1 Gyr, 1--2 Gyr and 12 -- 15 Gyr. The cluster formation episodes
identified in this study is in good agreement with the above results, but 
identified yet another episode of enhanced star cluster formation at 300 Myr.
Pietrzynski \& Udalski \cite{pu00} found peaks in star cluster formation
at 7 Myr, 125 Myr and 800 Myr. They also found peaks at 100 Myr and 160 Myr,
which they attribute to the last encounter of the Magellanic Clouds.

It is of interest to look for any correlation in star/cluster formation events
between the regions studied here, especially when the first four regions are not very far
from one another.
The regions 1 and 2 are separated by about 200 pc, and we see that the number
of clusters around these regions are very similar. The star formation has started a bit
early in region 2  and this region did not experience the 300 Myr episode of star
formation.  Therefore there is only very little similarity between the star formation
as well as cluster formation events around the regions 1 and 2, though the number of
star clusters found are similar.
Again, the number of clusters
identified near the regions 3 and 4 are very similar and the separation between
these two regions is also about 200 pc. 
The fraction of clusters formed at each age bin is also similar. The star formation
is found to have started a bit early in region 3. Both the two regions have experienced
the 300 Myr episode of star formation. Region 3 ended the star formation relatively
early. Therefore we see that the star formation and the cluster formation events in 
these two regions are well correlated. 
Therefore it is quite likely that the regions three and four underwent very similar
cluster and star formation events.
The separation between the left and the right extreme points of 
regions 3 and 4 are about  1 Kpc. We see that 
the cluster formation events are well corelated within a length scale of 1 Kpc.
This is similar to the length-scale found in the super giant shell LMC 4,
which is found to have a diameter of $\sim$ 1 Kpc (de Boer et al. \cite{b98}).
This might suggest that similar processes or cluster formation events have taken
place in the past in LMC.

 A look at the cluster fraction with respect to age, in
regions 5 and 6 reveal that these two regions have quite different 
age distribution of clusters. 
The region 5 is located well within the Bar and the
SFH is very different. This region seems to have experienced continuous 
star formation as well
as cluster formation, thoughout from 4 Gyr to 40 Myr. This probably is
due to the fact this is the central region, and likely to have had constant supply
of molecular clouds. The star formation occurring around this region, could also
have kept the star formation going on in this region. Also to be noticed is the
high density of stars found in this region, compared to the other regions studied here.

\subsection{Star formation events and LMC interactions}
We see from this study that three star formation events could be identified
from the CMDs of the field stars. 
And we find that the populous star clusters around the regions also follow
these three events. This trend is seen in all the 6 regions studied here.
One region (region 5) is found to have experienced continuous star 
formation. As it is located at the center of the bar, it is likely
that the star formation events in this region are unaffected by the factors influencing the
rest of the LMC.

The correlated star and cluster formation events which we see can be
due to some common triggering. These events could be due to some
external factors affecting the LMC. The most quoted and looked into mechanism for this,
is the interaction between LMC, SMC and the Galaxy. There are many studies which 
look for the signatures of possible encounters (Westerlund \cite{w97} and references 
therein, Maragoudaki \cite{m01}) and also studies 
which do simulation of the dynamics of
the interaction between the three galaxies (Fujimoto \& Murai \cite{fm84}, Gardiner \&
Neguchi \cite{gn96}, Gardinet et al. \cite{g94}). These have found that the LMC had
an interaction with our Galaxy at about 1.5 Gyr ago, 
with SMC at 0.2 -- 0.4 Gyr and LMC had a perigalacticon at 100 Myr. 
The close encounters between LMC and SMC were expected to
have occured at 1.5 Gyr, 3--4 Gyr, 5--6 Gyr, 7--8 Gyr and finally 10 Gyr ago (Vallenari \cite{v97}).

The star formation which started at 4 Gyr in LMC central regions does not seem to
have created star clusters. The 4 Gyr population is mainly seen inside the Bar region
of LMC and there are indications that the star formation started a little later in the 
outer regions of LMC, especially in the northern regions (see Subramaniam \& Anupama 
\cite{sa02}, Vallenari \cite{v97}). 
 The star formation, even in the outskirts of the Bar is found to have started 
a little later, like 3.2 Gyr, 2.5 Gyr etc.. 
Probably the 4 Gyr star formation is mainly associated with the Bar.
It will be interesting to find out whether the 4 Gyr star formation
episode corresponds to the formation of the Bar in LMC.
The populous star clusters are formed somewhere between 1 and 3 Gyr. One does not
know whether this cluster formation event has anything to do with the LMC-SMC-Galaxy
interaction at about 1.5 Gyr. 

The 300 Myr cluster formation
episode is seen in the whole of LMC and also in SMC (Maragoudaki et al. \cite{m01}),
confirming that the both the galaxies are affected. It is quite clear that the star
formation and the cluster formation events in LMC and SMC were started due their
interaction. Therefore, it is quite likely that the populous star clusters formed at this 
time are as a result of star formation mechanisms due to interactions.
Some clusters could also have been formed due to the propagating star
formation started by the triggers. 
Similarly, the star and the cluster formation episodes seen around 100 Myr could be due
to the perigalacticon passage of LMC. This could have triggered the star formation events,
leading to the formation of populous star clusters. Probably this event 
triggered star formation in SMC as well.  To summarise, this study shows that, among
 the three episodes
of correlated star and cluster formation events, two seems to have started due to
the interaction of LMC with Galaxy or SMC. A proper understanding of the dynamical history
of the LMC, SMC and Galaxy will probably shed some light on the possible reasons for the
4 \,--\, 2.5 Gyr star and cluster formation episodes.

\section{Conclusions}
The present study focussed on the correlaton between star and cluster formation in six regions
of the LMC. The results are summarised blow:
\begin{itemize}
\item[1.] The star formation around few tens of parsec and cluster formation around 400 pc
centered at six regions are studied. It is seen that in all the six regions, the star formation
events and the cluster formation events are more or less correlated. Though coresponding to
the youngest clusters, aged less than 30 Myr, no stellar population is seen.
\item[2.] The cluster formation events and the fraction of star clusters formed at these events
were found to be very similar for the two regions located to the south-east of the Bar.
This might suggest that these regions experienced similar cluster formation triggers at almost
similar instances.
\item[3.] The major star formatoin events are found to be three in most of the regions studied here.
The star formation which began about 4 Gyr ago, continued upto around 1 Gyr, or continued further.
The other two events have taken place around 300 Myr, and at 100 Myr.
\item[4.] Two of the above three events seem to  correlate with the 
interactions of LMC with
SMC and Galaxy, except the 4 Gyr event. The cluster and the star formation events at 300 Myr and 
100 Myr are likely to be triggered by the galaxy interactions. Therefore, the populous star
clusters formed at these instances are likely to be the product of star formation mechanisms due
to triggers arising from interactions and some from the propagation of these events.
\end{itemize}

\end{document}